\documentclass[pra,twocolumn,showpacs]{revtex4}
\usepackage{times}
\usepackage{graphicx}
\usepackage{color}
\newcommand{\be}{\begin{equation}}
\newcommand{\ee}{\end{equation}}
\newcommand{\surfCharge}{\gamma}
\newcommand{\epsZero}{\epsilon_\infty}

\newcommand{\vCP}{vdW-CP }
\begin{document}

\title{Van der Waals--Casimir--Polder interaction of an atom with a 
composite surface}
%\subtitle{}
\author{Elad Eizner$^{1}$, Baruch Horovitz$^{1}$,
and Carsten Henkel$^{2}$
% \thanks{\emph{Present address:} Insert the address here if needed}%
}                     % Do not remove
\affiliation{
$^{1}$Physics Department,
Ben Gurion University of the Negev,
Beer Sheva,
Israel\\
$^{2}$Institut f\"ur Physik und Astronomie,
Universit\"at Potsdam,
Germany}
\date{Received: date / Revised version: date}
\begin{abstract}
We study the dispersion interaction of the van der Waals and Casimir--Polder 
(vdW-CP) type between a neutral atom and the surface of a metal by allowing for nonlocal electrodynamics, i.e. electron diffusion. 
We consider two models: (i) bulk diffusion, and (ii) diffusion in a surface charge layer. In both cases the transition to a semiconductor is continuous 
as a function of the conductivity, unlike the case of a local model. The relevant 
parameter is the electric
screening length and depends on the carrier diffusion constant. 
We find that for distances comparable to the screening length,
\vCP data can distinguish between bulk and surface diffusion, 
hence it can be a sensitive probe for surface states.
\end{abstract} %end of abstract
%\journalname{The European Physical Journal D}
\maketitle

\section{Introduction}
\label{s:intro}

Recent progress in the understanding of the van der Waals--Casimir--Polder 
(vdW-CP) force between
an atom and a surface allows by now to distinguish surface properties with
respect to charge transport. Data on the temperature and atom-surface distance dependence provide excellent tools for such analysis. In particular the experiments on fused silica \cite{harber} have demonstrated a temperature dependence of the \vCP interaction. Fused silica is considered as a dilutely doped semiconductor that has a finite conductivity $\sigma$. The data was fitted successfully to the potential for a dielectric surface 
which differs from the one for a perfectly reflecting mirror, as considered by Casimir and Polder \cite{cp}. In fact, since $\sigma/\omega$ diverges at zero frequency, any nonzero conductivity will reproduce the perfect reflector 
result \cite{Klimchitskaya08}.
% CH 19 Jun 12 sentence changed
% 
% , although according to the original prediction of Casimir and Polder \cite
%{cp,Klimchitskaya08},
% the fact that $\sigma/\omega$ diverges at zero frequency
% %even for low conductivities, 
% leads to a different result. 
%
% This puzzle has been resolved with the help of a 
% nonlocal model of the electromagnetic response, where charge diffusion 
% and screening become essential 
%
% at low conductances \cite{Pitaevskii08}. 
%
% CH 19 Jun 12 sentence changed
It has been suggested to resolve this puzzle with the help of a nonlocal model of the electromagnetic response, where charge diffusion and screening become essential at low conductances \cite{Pitaevskii08}, although the approach was met with criticism \cite{comments-and-reply}.
In a related experiment on the (macroscopic)
Casimir force between a gold-coated sphere and a single-crystal silicon membrane \cite{Chen2007}, the results are consistent with a dielectric behavior in its pristine form (ignoring the $\sigma/\omega$ tail). The significant change in the charge carrier density after laser illumination leads to a metallic response in local form, 
again in
agreement with Casimir force measurements. 
% CH 19 Jun 12 -- sentence changed
% The crossover between these limits is now understood in terms of a nonlocal 
% theory as well \cite{Pitaevskii08,Dalvit08}.
%
Nonlocal theories have been worked out to understand the crossover between these limits \cite{Pitaevskii08,Dalvit08,Svetovoy2008}, although experimental data favor a local description 
% CH 29 Jun 12 -- half-sentence changed
where the contribution of free charge carriers is omitted in the dielectric state
\cite{Chen2007,comments-and-reply}. For the discussion whether
nonlocal electrodynamics may be applied to macroscopic Casimir interactions,
and its consistency with thermodynamics, see Refs.\cite{Barash75,comments-and-reply,%
comments-and-reply-Dalvit08,KlimchitskayaJPhysA2008,%
PitaevskiiLaserPhysics2009}.

The ability of probing charge transport may provide for an increased understanding of surface science by using the \vCP interaction as a probe. 
This viewpoint may be traced back to the seminal paper by Zaremba and
Kohn \cite{Zaremba1976} where the van der Waals potential (neglecting
retardation) was calculated
with a microscopic description of the many-electron response of a metal.
Their analysis yields an expression for the reference
plane with respect to which the atom-surface distance $z$ is actually 
calculated. See Ref.\cite{Dobson2012} for a review of related methods.
Dorofeyev \cite{Dorofeyev2007} analyzed
the van der Waals (non-retarded) regime with the help of
a nonlocal ($k$-dependent) electromagnetic response based on the 
surface impedance work by Kliewer and Fuchs~\cite{Kliewer1968};
the electrons were assumed to reflect specularly from the inner surface.
%
% {\color{magenta}
% $\bullet$ nonlocal effects in the \vCP literature
% }

In the present work, we compare two non-local models that can be understood
as mesoscopic extensions of the work by Zaremba and Kohn: the first one
allows for bulk diffusion as in Refs.\cite{Pitaevskii08,Dalvit08}, we denote it
``continuous charge'' 
(CC). In the non-retarded limit, this reduces to the Kliewer and Fuchs approach
for a hydrodynamic dielectric function in the bulk. The model (ii) allows for 
diffusion only within a surface layer (``charge layer'' or CL), 
the bulk charges responding with a local conductivity. Such composite surfaces with charges in the bulk and on the surface, are fairly common in metallic systems. These surfaces are either covered with adsorbates or nanostructures and can, e.g., be used as sensitive chemical sensors and biosensors \cite{homola}, or are disordered with quantum well states at the surface \cite{varykhalov}. Further motivation for a two-type charge model comes from studies of
the anomalous heating of cold ions observed in miniaturized Paul traps
that invoke surface charge fluctuations on the metallic
electrodes \cite{Turchette00a,Leibrandt07b,Henkel08,Dubessy09,Daniilidis11}. Composite surfaces have also been explored regarding surface plasmons
and lead to a wide range of dispersion relations, as 
observed in different systems, see Ref.\cite{Horovitz2012} and references
therein.

The hallmark of the non-local theory is charge diffusion. 
In the CC model, it is described by the diffusion coefficient $D$ in the bulk, while
$D_s$ describes surface diffusion in the CL. We identify a screening 
length $a_0$ for CC (cgs units)
\begin{equation}
	a_0 = \sqrt{ \frac{ \epsZero D }{ 4\pi \sigma } }
,	\label{eq:def-a0}
\end{equation}
where $\sigma,\epsZero$ are the bulk conductivity and background dielectric constant, respectively (replace $D$ with $D_s$ in the CL model).
% the 
% $a_0=\sqrt{\epsZero D_s/4\pi \sigma}$.
%and in the thermal range of the CP effect, i.e. $z_0$ larger than the wavelength of the dipole transition.
As the atom-surface distance $z$ becomes comparable to the length scale $a_0$,
the \vCP interaction changes its behaviour. We consider different materials
where $a_0$ can be compared with the other two 
important length scales of the \vCP potential:
the radiation wavelength $\lambda_A = c/\Omega$, $\Omega$ being the 
atomic transition frequency, and the thermal photon wavelength $\lambda_T
= \hbar c/k_BT$ at temperature $T$. As common for electric dipole transitions 
in the visible range, we assume here 
% that $k_BT\ll \hbar\Omega$,  hence 
$\lambda_A\ll\lambda_T$, but explore otherwise the full range of distances.
% The screening length $a_0$ adds a new important length scale. 
After defining the charge models (section~\ref{s:model}), we study in section~\ref{s:van-der-Waals} the case that $a_0$ is in the van der Waals (short distance) regime, $a_0\ll\lambda_A$, focusing on the $T=0$ case. In section~\ref{s:Casimir-Polder},
the length $a_0$ is in the Casimir-Polder (intermediate) regime,
$\lambda_A\ll a_0\ll \lambda_T$, while in section~\ref{s:zero-freq-term} we consider long distances $a_0\gg \lambda_T$ where screening affects the Lifshitz (thermal) regime of the interaction. In all these cases we find that the crossover from the \vCP potential for a local conductor to a dielectric occurs at $z\approx a_0$. In addition, the dispersion interaction is a good surface probe in the sense that the crossover is different in the CC and CL systems. This difference is particularly visible when the two limits, metallic and dielectric, are well separated. The relevant parameters are discussed in the conclusion.

\section{Model}
\label{s:model}

\subsection{Atom--surface interaction potentials}
\label{s:Wylie-and-Sipe-approach}

We use in this paper the general formulation of Wylie and Sipe
\cite{Wylie1984}
for the Casimir-Polder potential of an atom with a surface.
Assuming the surface and the electromagnetic field in equilibrium
at temperature $T$, the free energy of interaction is given by
\begin{equation}
{\mathcal{F}}( {\bf r} ) = - k_{\rm B} T
	\sum_{l=0}^{\infty}\kern-0.1em\raisebox{0.7ex}{$'$}
	\sum_{ij}
	\alpha_{ij}(i\xi_l)G_{ji}({\bf r}, {\bf r}; i\xi_l)
	\label{eq:F-CP-by-Wylie-and-Sipe}
\end{equation}
which is a sum over Matsubara frequencies 
$\xi_l \equiv 2\pi l k_{\rm B} T / \hbar$ along the imaginary frequency
axis. The primed sum is taking the $l=0$ term with a factor $\frac{1}{2}$.
The atomic polarizability $\alpha_{ij}( \omega )$ is given in 
Eq.(\ref{eq:def-alpha}) below,
the retarded Green tensor $G_{ij}({\bf r}, {\bf r}; \omega )$ is
made explicit in Eqs.(\ref{eq:Gxx-integral},\ref{eq:Gzz-integral}).
In the limit $T\rightarrow 0$, ${\cal F}$ reduces to the interaction energy
\begin{equation}
U( {\bf r} ) = - \frac{\hbar}{2\pi}\int^{\infty}_{0}\!{\rm d} \xi
\sum_{ij} \alpha_{ij}(i\xi)
G_{ji}({\bf r}, {\bf r}; i\xi)
	\label{eq:U-CP-zero-temp}
\end{equation}
% Assuming $k_{\rm B} T \ll \hbar \Omega$ (see Introduction), 
This formulation applies for an atom in the ground state, for which the
polarizability tensor is given by
\begin{equation}
	\alpha_{ij}(\omega) = \lim_{\eta\rightarrow 0^+} \frac{2}{\hbar}
	\sum_e
	\frac{\Omega_{eg} d_{i}^{ge}d_{j}^{eg} 
		}{ 
	\Omega_{eg}^2 - (\omega+i\eta)^2}
	\label{eq:def-alpha}
\end{equation}
where $\Omega_{eg}$ is the transition frequency between the ground state 
($g$) and an excited state
($e$) with an electric dipole matrix elements
$g \leftrightarrow e$ and $d_i^{ge} = \langle g | d_i | e \rangle$. We 
focus on a single resonance and assume rotational symmetry, so that 
$\Omega_{eg} = \Omega$ and the polarizability
is isotropic, $\alpha_{ij} = \alpha \,\delta_{ij}$.

The electromagnetic Green tensor $G_{ij}( {\bf r}, {\bf r}'; \omega )$
in Eq.(\ref{eq:F-CP-by-Wylie-and-Sipe})
provides the electric field $E_i( {\bf r} )$ radiated by a test dipole
located at ${\bf r}'$ and oscillating with amplitude $d_j$ at frequency
$\omega$:
\begin{equation}
	E_i( {\bf r} ) = {\textstyle\sum_{j}}
	G_{ij}( {\bf r}, {\bf r}'; \omega ) d_j
	\label{eq:def-Green}
\end{equation}
For a source outside a polarizable body, this electric field can be calculated
within macroscopic electrodynamics (see, e.g., Ref.\cite{Wylie1984}) and
involves the reflection (or scattering) amplitudes of the body. 
It turns out that these amplitudes are sufficient to determine
the Casimir-Polder interaction. The subtraction of the free-space part of the 
Green tensor (Lamb shift) is understood in the following.
At a planar surface, only two principal polarizations
$p = {\rm TM}, {\rm TE}$ are relevant, and the reflection amplitudes depend
on frequency and a wave vector $k$ parallel to the surface
$r_p = r_p( \omega, k )$. % ($p = {\rm TM}, {\rm TE}$).
As we put ${\bf r} = {\bf r}'$ in Eqs.(\ref{eq:F-CP-by-Wylie-and-Sipe}, 
\ref{eq:U-CP-zero-temp}), the planar symmetry implies
that the Green tensor is diagonal with elements \cite{Wylie1984}
(cgs units)
\begin{eqnarray}
	G_{xx}( {\bf r}, {\bf r}; \omega ) &=&
	\int_0^\infty \! {\rm d}k\, \frac{ k e^{-2 v_0 z}  }{ 2v_0 }
	\left[ v_0^2 r_{\rm TM}% ( \omega, k )
	+
	\frac{\omega^2}{c^2}r_{\rm TE}%( \omega, k )
	\right]
	\label{eq:Gxx-integral}
\\
	G_{zz}( {\bf r}, {\bf r}; \omega ) &=&
	\int_0^\infty \! {\rm d}k\, \frac{ k^3 e^{-2 v_0 z}  }{ v_0 }
	r_{\rm TM}%( \omega, k )
	,
	\label{eq:Gzz-integral}
\end{eqnarray}
the element $G_{yy}$ being identical to $G_{xx}$.
%
% conjecture for free-space Green tensor, imaginary part:
% take r_TE piece in G_xx, set z = 0, integrate over 0 < k < \omega / c.
% multiply with 2 for the other polarization.
%
% this gives 
%\begin{eqnarray}
%	G_{xx}^{\rm fs} &=& \frac{\omega^2}{c^2}
%	\int_0^{\omega/c} \! {\rm d}k\, 
%	\frac{ i k }{ \sqrt{ \omega^2 / c^2 - k^2 } }
%\nonumber\\
%	&=& \frac{ i \omega^3 }{ c^3 }
%	\label{eq:free-space-Gxx}
%\end{eqnarray}
% this gives $G_{xx} = 1\,{\rm cm}^{-3}$ at $\omega = 
% 3 \cdot 10^{10}\,{\rm rad}/{\rm s}$, seems to conform with the plot.
%
% The arguments $(\omega, k )$ have been suppressed for brevity.
The vacuum decay constant for a field mode with frequency $\omega$
and parallel wave vector $k$ is
 \begin{equation}
	v_0 = \sqrt{k^2 - \omega^2/ c^2}
	\label{eq:def-v-0}
\end{equation}
with the root chosen such that
${\rm Re} \, v_0 \ge 0$ and ${\rm Im}\,v_0 \le 0$.
The reflection amplitudes are collected in Table~\ref{tab:reflection coefficients}
for the different surface models considered in this paper.
%By the symmetry of the planar surface, the other elements of $G_{ij}(
%{\bf r}, {\bf r}; \omega )$ vanish.
% We use cgs units here; in SI units, multiply $G_{ij}$ with the factor
% $1/(4\pi\varepsilon_0)$.
%
The general Eq.(\ref{eq:F-CP-by-Wylie-and-Sipe}), valid for any $T$, 
now takes the form
\begin{eqnarray}
&&\mathcal{F}( z ) = -\frac{k_{\rm B}T}{2}\int_0^\infty\! {\rm d}k \,
2k^2e^{-2kz}\alpha(0)r_{\rm TM}(0,k) 
	\nonumber\\
&& {} - k_{\rm B} T\sum_{l=1}^\infty \alpha(i\xi_l)
\int_0^\infty\! {\rm d}k \,
\frac{k}{v_0}e^{-2 v_0 z}
\Big[ 2 k^2 r_{\rm TM}(i\xi_l,k) + {}
\nonumber\\
&& {} + \frac{\xi_l ^2}{c^2} ( r_{\rm TM}(i\xi_l,k) - r_{\rm TE}(i\xi_l,k) )
\Big]
	\label{eq:Wylie-Sipe-finite-T}
\end{eqnarray}
% {\color{magenta}discussion on discontinuity shifted to thermal range section}

Let us briefly recall the assumptions behind the Wylie and Sipe approach
\cite{Wylie1984}:
the interaction energy is calculated in perturbation theory with respect to
the atom-field coupling, starting from a well-defined atomic level (here,
the ground state). The temperature provides Boltzmann weights for the
excited states of the electromagnetic field, thus including the interaction
with blackbody radiation and its modification by the surface. The thermal
population of excited states of the atom is negligible provided the Bohr
frequency is large enough,
$\hbar \Omega \gg k_{\rm B} T$.
Otherwise, a temperature-dependent polarizability should be
used in Eqs.(\ref{eq:F-CP-by-Wylie-and-Sipe}, \ref{eq:Wylie-Sipe-finite-T}). 
Finally, the surface response
is worked out ignoring the presence of the atom and assuming a linear
response of the surface to electromagnetic radiation, consistent with common
practice in surface spectroscopy. 
By inspection of
Eq.(\ref{eq:Wylie-Sipe-finite-T}), one notes that wave vectors
up to $k \sim 1/z$ are relevant for the interaction potential. At distances
$z$ much larger than the size of the unit cell, a macroscopic treatment 
of the surface response is therefore justified.
%
% CH 28 Jun 2012 sentence added
There have been discussions what kind of electromagnetic response may be used 
consistently for atom-surface potentials and macroscopic Casimir interactions 
in general, see Refs.\cite{Barash75,comments-and-reply,%
comments-and-reply-Dalvit08,KlimchitskayaJPhysA2008,%
PitaevskiiLaserPhysics2009}. The approach of Ref.\cite{Wylie1984} is based
on the fluctuation-dissipation theorem \cite{Landau9}.

\subsection{Surface response with charge diffusion}

\begin{table*}%[t]% use {table*} if two columns wide
%\begin{center}
\begin{tabular}{| llll |}
\hline
%\noalign{\smallskip}
\rule[-1.5ex]{0pt}{4ex}%
& local & hydrodynamic bulk charge (CC) & charge layer (CL)  
\\
%\noalign{\smallskip}
\hline
%\noalign{\smallskip}
\rule[-2.5ex]{0pt}{6ex}%
$r_{\rm TE}(\omega,k)$
	& $\displaystyle \frac{v_0-v}{v_0+v}$
	& $\displaystyle \frac{v_0-v}{v_0+v}$
	& $\displaystyle \frac{v_0-v}{v_0+v}$
\\
\rule[-3.5ex]{0pt}{7ex}%
$r_{\rm TM}(\omega,k)$
	& $\displaystyle \frac{\epsilon v_0 -v}{\epsilon v_0 +v}$
	& $\displaystyle \frac{\epsilon v_0 - v - (\epsilon-\epsZero)\frac{k^2}{\epsZero v_1}}{\epsilon v_0 + v +(\epsilon-\epsZero)\frac{k^2}{\epsZero v_1}}$
	& $\displaystyle \frac{\epsilon_s v_0 -v}{\epsilon_s v_0 +v}$
\\
%\noalign{\smallskip}
\hline
\end{tabular}
%\end{center}
% Or use
%\vspace*{5cm}  % with the correct table height
%\vspace*{0.5ex}
%
\caption{The reflection coefficients $r_{\rm TE}(\omega,k)$ and $r_{\rm TM}(\omega,k)$ for two models for a nonlocal surface response.
\\
Symbols used: The dielectric function $\epsilon( \omega )$ is defined at (\ref{local_dielectric}) while
$\epsilon_s(\omega, k)$ is defined at (\ref{eq:CL_dielectric}).
Spatial decay rates are for the vacuum $v_0( \omega, k )$,
Eq.(\ref{eq:def-v-0}), for transverse fields in the medium 
$v(\omega, k) = \sqrt{k^2 - \epsilon(\omega)\omega^2 / c^2}$
with ${\rm Re}\,v > 0$ and for compressional charge waves
$v_1(\omega, k ) =
\sqrt{ k^2 - i \omega \epsilon(\omega) / [\epsZero D(\omega)] }$
with ${\rm Re}\,v_1 > 0$ and $D(\omega) = D / (1-i\omega\tau)$.
}
\label{tab:reflection coefficients}       % Give a unique label
\end{table*}% use {table*} if two columns wide

\begin{figure*}[hbt]
\includegraphics*[width=0.4\textwidth]{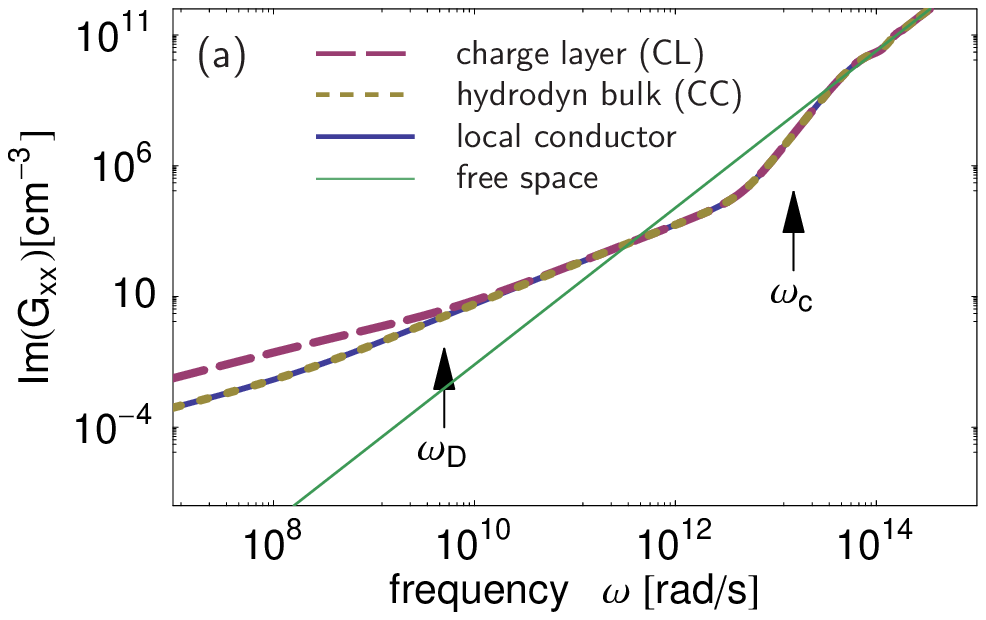}
\hspace*{0.1in}
\includegraphics*[width=0.4\textwidth]{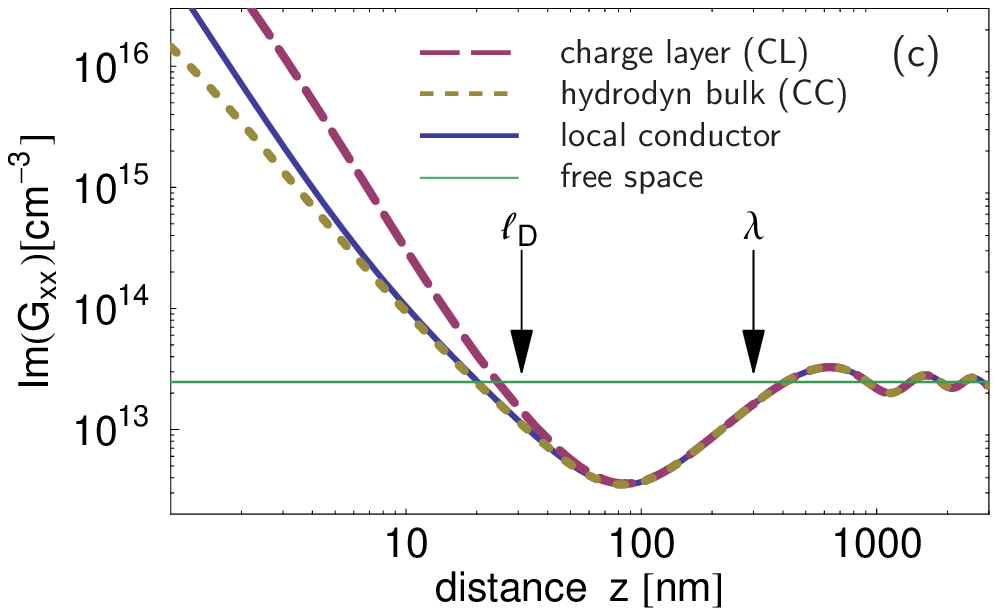}

\includegraphics*[width=0.4\textwidth]{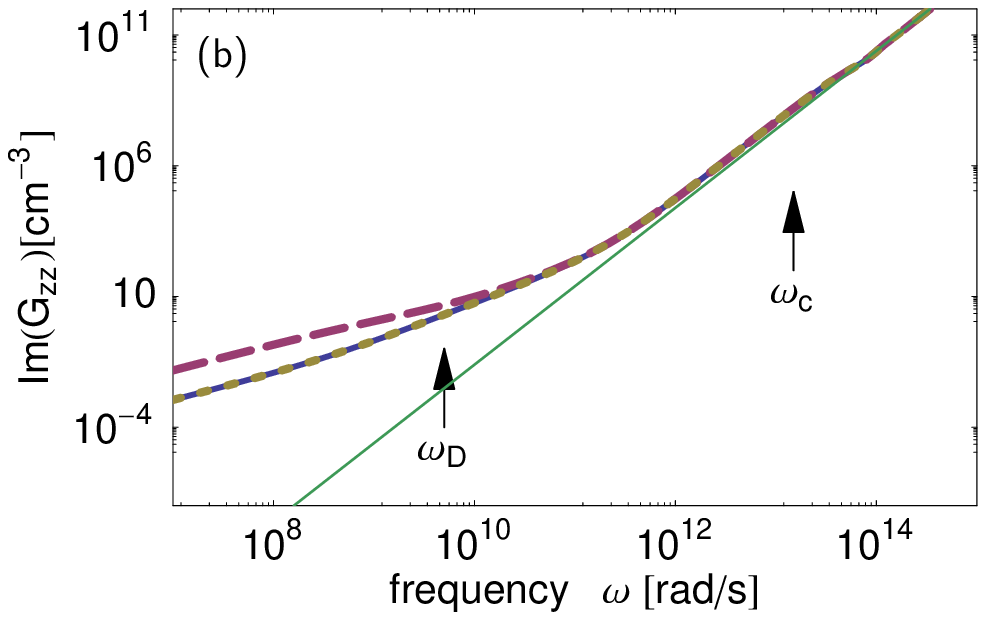}
\hspace*{0.1in}
\includegraphics*[width=0.4\textwidth]{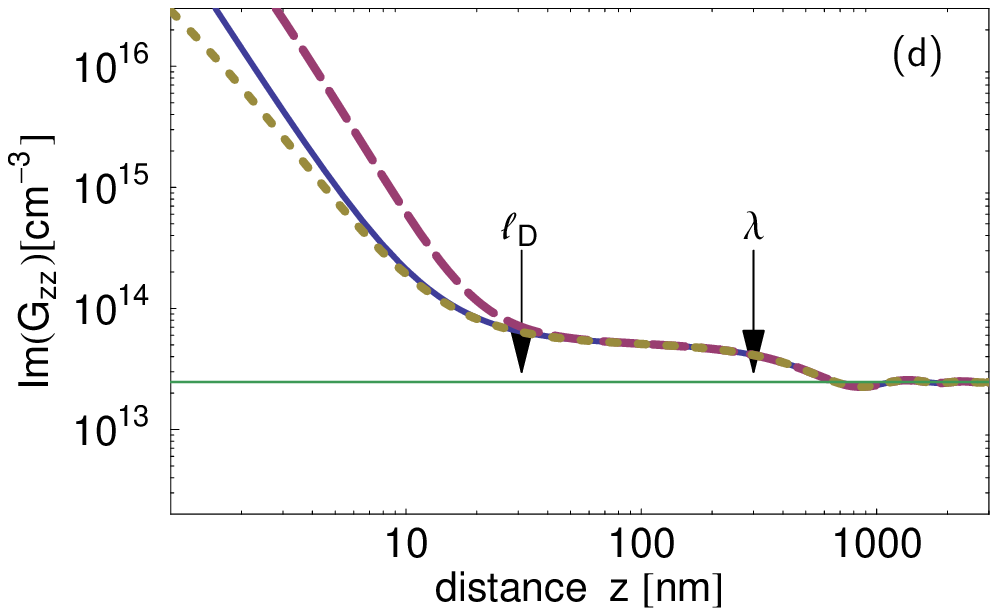}

% if figure file is missing
\caption[]{(\emph{left column})
Local photonic mode density near a metallic surface,
described as a local Ohmic conductor (solid blue curves),
by a continuous bulk charge (CC) model (dotted yellow curves,
superimposed),
and a diffusive charge layer (CL) model (dashed purple curves). 
Thin green lines: free space mode density.
We plot the imaginary part of
the Green functions in Eqs.(\ref{eq:Gxx-integral}, \ref{eq:Gzz-integral});
top row: polarization parallel to surface ($G_{xx}$), bottom row:
perpendicular polarization ($G_{zz}$).
Parameters:  conductivity
% 4\pi\epsZero \sigma_{cgs} = 4 \cdot 10^{7} / (\Ohm m)
% hence specific resistance 2.5 10^{-8} \Omega m, consistent with Al
$\sigma = 3.6 \cdot 10^{17}\,{\rm rad}/{\rm s}$, typical for Al,
relaxation time $\tau = \tau_s = 10^{-15}\,{\rm s}$,
diffusion constant $D_s = D = 5 \cdot 10^3\,{\rm cm^2}/{\rm s}$,
background permittivity $\epsZero = 1$, giving a screening length
$a_0 \approx 0.3\,{\rm nm}$, and
distance $z = 10 \,\mu {\rm m}$. The plots include the free space
response that provides the dominant $\sim \omega^3$ scaling at high
frequencies.
Characteristic diffusion frequency for these parameters:
$\omega_D = D / z^2 = 5 \cdot 10^{9} \,{\rm rad}/{\rm s}$ (left
arrow); right arrow: inverse photon round trip time $\omega_c =
c / ( 2 z )$.
\\
(\emph{right column}) Distance dependence of the local mode density
near a metallic surface,
described by the same models as in the left column.
Same parameters, except that the frequency is fixed to
%$\sigma=3.6\cdot 10^{17}[\frac{1}{s}], \tau=\tau_s=10^{-15}[s],
$\omega = 10^{15}\,{\rm rad}/{\rm s}$
(near-infrared).
% wavelength 2\pi c / \omega = 1.88\,\mu{\rm m}, quite large
%  D=D_s=5\cdot10^3[\frac{cm^2}{s}]$,
The plots include the free space response, that leads to a constant limit
$\frac23 \omega^3$ at large distance (thin horizontal lines).
The oscillations are due to partial standing waves
formed above the surface.
Diffusion length for these parameters:
$\ell_D = \sqrt{ 2 D / \omega } \approx 31 \,{\rm nm}$ (left arrow);
right arrow: reduced wavelength $\lambda = c / \omega$.
}
\label{fig:LDOS-metal}
% \label{fig:LDOS-metal2}
\end{figure*}

Surfaces covered with thin layers of strongly localized charges or adsorbates
have been studied in much detail in surface physics. For a general theory
of their electromagnetic response, see Refs.\cite{Sipe80,BedeauxVlieger}.
We consider here
a model introduced in Ref.\cite{Henkel08} where the surface is covered
by a charge sheet with a charge density $\gamma(x,y)$
(localized in the plane $z = 0$).
The details of the electromagnetic response
are worked out in Appendix~\ref{a:calculation-TE-TM}. The sheet current 
in the layer responds by diffusion
\begin{equation}
	{\bf J} = - D_s( \omega ) \nabla_\Vert \surfCharge
	\label{eq:diffusive-current-response}
\end{equation}
where $D_s(\omega) = D_s / (1 - i \omega \tau_s)$ is the 
surface diffusion coefficient and $\tau_s$ a surface relaxation 
time. The gradient appearing here is parallel to the layer.
A surface conductivity term proportional to ${\bf E}$ 
is neglected, as justified for 
a small layer thickness, see Appendix~\ref{a:no-surface-conduction}.
Charge conservation yields
\begin{equation}
	- {\rm i} \omega \surfCharge + \nabla \cdot {\bf J} =
	j_z( 0^- )
	\label{eq:layer-bulk-charge-conservation}
\end{equation}
so that the bulk current just below the layer, $j_z( 0^- )$, provides the
influx into the surface layer. We take the bulk current response in the usual
Ohmic form
\begin{equation}
z < 0: \quad
{\bf j} = \sigma( \omega ) {\bf E}
	\label{eq:Drude-metal}
\end{equation}
where the Drude conductivity is $\sigma( \omega ) = \sigma /
( 1 - {\rm i} \omega  \tau )$ with a scattering time $\tau$. Eqs. (\ref{eq:diffusive-current-response},\ref{eq:Drude-metal}) define the CL (charge
layer) model. 

The CC (continuous charge) model, in contrast, is defined by the bulk charge
density 
that has a diffusion constant $D$, i.e. Eq. (\ref{eq:Drude-metal}) is replaced by
\begin{equation}
{\bf j} = \sigma( \omega ) {\bf E} - D( \omega ) \nabla \rho
	\label{eq:hydrodynamic-conductivity}
\end{equation}
The (``additional'') boundary condition for the current is 
then $j_z( 0^- ) = 0$,
since there are no surface charges.

% table and figure shifted above

The resulting reflection amplitudes are summarized in Table~\ref{tab:reflection coefficients} and more details on their derivation
are given in Appendix A.
One notes that the TE polarization is not affected by the composite
structure of the surface. This can be understood from the fact that
surface charges are created by electric fields perpendicular to the
surface, which are absent in this polarization.

To illustrate the impact of the diffusive layer, we have calculated the
local photonic mode density, i.e., the imaginary part of
$G_{ii}( {\bf r}, {\bf r}; \omega )$. This quantity can be measured
from the spontaneous decay rate of an excited atom placed at ${\bf r}$
or from the heating rate of an ion trapped near a surface
\cite{Turchette00a,Leibrandt07b,Henkel08,Dubessy09,Daniilidis11}.
The results shown in Fig.\ref{fig:LDOS-metal}
illustrate the enhancement of the mode density
at low frequencies (below the characteristic scale $D_{(s)} / z^2$
left column). At large wave vectors (short distances, right column),
there is a competition
between additional modes (enhancing the mode density, CL model) and 
screening (reducing it, CC model). Note that for the parameters considered 
here the screening length $a_0$ is much smaller than the diffusion
length $\sim \sqrt{ D_{(s)} / \omega }$). An excited atom decays faster
because diffusion along the surface broadens the field spot it creates, increasing 
the effective area where absorption takes place.
Calculations of the Casimir (plate-plate) interaction between materials
with a nonlocal electromagnetic response have revealed qualitatively
similar trends (compare Refs.\cite{Esquivel04a,Svetovoy06a} to
Ref.\cite{Contreras-Reyes05a}).
%
% CH 28 Jun 12, sentence added.
The experimental data of Ref.\cite{Chen2007} are better described with a local
rather than nonlocal theory, however, see Ref.\cite{KlimchitskayaJPhysA2008}
and the discussion in Refs.\cite{MostepanenkoIJMPA2009,PitaevskiiLaserPhysics2009}.
%
% {\color{green}Check these formulations, as in the vdW-CP results, we
% rather see a suppression of interaction (key word ``screening'').}

%Discussion for metal: make a plot for real frequency
%(around a typical atomic absorption line) of the local photonic mode density
%${\rm Im}\,G_{ii}( {\bf r}, {\bf r}; \omega )$. Fixed distance, fixed frequency.
%
%Compare with and without charge layer. Local calculation and hydrodynamic
%model (with specular boundary condition).

%\cleardoublepage

%
\section{Van der Waals (nonretarded) regime}
\label{s:van-der-Waals}

This regime corresponds to short distances where retardation is negligible
\begin{equation}
	z \ll \lambda_A \equiv c / \Omega
	\label{eq:van-der-Waals-limit}
\end{equation}
with a typical value $\lambda_A\approx 100\,{\rm nm}$ for transitions in 
the visible range. 
The Van der Waals interaction follows a power law 
${\cal F}( z ) \sim 1/z^3$
for a material with a local response (Drude metal or dielectric). 
We consider in this section the situation that the screening length
(see Introduction) satisfies $a_0 \ll \lambda_A$;
this corresponds to electron densities typical for metals.

% We may use therefore high conductivity
% $\sigma\gg \Omega$, though this is not assumed below. 
We start from the zero-temperature expression for the interaction
potential, 
combining Eqs.(\ref{eq:U-CP-zero-temp}, \ref{eq:def-alpha},
\ref{eq:Gxx-integral}, \ref{eq:Gzz-integral}):
\begin{eqnarray}
U( z ) &=& -\frac{\hbar}{2\pi}\int^{\infty}_{0} \!{\rm d} \xi \, \alpha(i\xi)
\int^{\infty}_{0} \!{\rm d}k \frac{k}{v_0}e^{- 2 v_0 z}\cdot {}
	\label{eq:CP-zero-T-imag-freq}
\\
&& {} \cdot \left[
2 k^2 r_{\rm TM}( {\rm i}\xi, k )
+ (\xi ^2/ c^2) ( r_{\rm TM}( {\rm i}\xi, k ) - r_{\rm TE}( {\rm i}\xi, k ) )
\right]
\nonumber
\end{eqnarray}
where now $v_0^2 = k^2 + \xi^2 / c^2$.
The dominant ranges of the integrals are
around $k \sim 1 / z \gg \Omega / c$,
due to the exponential, and 
$\xi \leq \Omega$, due to the polarizability $\alpha( {\rm i}\xi )$.
This allows to simplify
Eq. (\ref{eq:CP-zero-T-imag-freq}) by taking $\xi \ll c k$ and $v_0 \approx
k$, so that for the CL model (see Table~\ref{tab:reflection coefficients})
\begin{eqnarray}
	U( z ) &\approx& - \frac{\hbar \alpha(0) }{ \pi }
	\int^{\infty}_{0}\! {\rm d}\xi \,
	\frac{ \Omega^2 }{ \xi^2 + \Omega^2 }
\cdot {} \nonumber\\
	&&
{} \cdot
	\int^{\infty}_{0}\! {\rm d}k \, e^{-2 k z} k^2
\frac{\epsilon_s( {\rm i}\xi, k ) k - v}{\epsilon_s( {\rm i}\xi, k ) k+ v}
	\label{eq:vdW-integral-xi-CL}
\end{eqnarray}
The reflection amplitude involves a ``surface dielectric function'' 
given by (see Ref.\cite{Henkel08}
and Appendix~\ref{a:calculation-TE-TM})
\be
%\epsilon_s(\omega, k) = \frac{
%\epsilon(\omega) + i\epsZero D_s(\omega) k^2 / \omega
%}{ 1 + iD_s(\omega) k^2 / \omega}
\epsilon_s(\omega, k) = \epsZero + 
\frac{
	4\pi i \sigma
}{ (\omega + iD_s(\omega) k^2 ) (1 - {\rm i} \omega \tau) }
	\label{eq:CL_dielectric}
\ee
where $D_s( \omega )$ is the surface diffusion coefficient 
[see Eq.(\ref{eq:diffusive-current-response})].
% where $D_s(\omega)=D_s/(1-i\omega\tau_s)$ includes a surface 
% relaxation time. 
The conventional Drude dielectric function (local), with a bulk 
relaxation time $\tau$ and a high-frequency asymptote $\epsZero$, is
\be
\epsilon( \omega ) =
\epsZero + \frac{4\pi i \sigma}{ \omega (1 - {\rm i} \omega \tau)}
\label{local_dielectric}
\ee
which happens to be the limiting form of Eq.(\ref{eq:CL_dielectric})
when $D_s \to 0$. From the $k$-dependence 
in $\epsilon_s(\omega, k)$, 
we identify the dimensionless ratio
$z^2 4\pi \sigma/(\epsZero D_s) = (z/a_0)^2$ that defines 
the screening length $a_0$ % \equiv \sqrt{\epsZero D_s/4\pi \sigma}$,
consistent with the estimate~(\ref{eq:def-a0}) of the Introduction.
For $k a_0\ll 1$, the reflection amplitude in Eq.(\ref{eq:vdW-integral-xi-CL})
recovers the local behaviour, while at very short distances,
$k a_0\gg 1$, the diffusive (nonlocal) term dominates. The latter case 
implies that the conductivity contribution is suppressed in
Eq.(\ref{eq:CL_dielectric}), leaving only the
background dielectric constant $\epsZero$.

If $\epsZero > 1$, both limiting cases (conductor and dielectric) show
a van der Waals interaction that follows the familiar $c_3/z^3$ power law,
but with different $c_3$ coefficients. 
For the dielectric, 
\begin{equation}
	\mbox{local diel.}, z \ll \lambda_A : \quad
	U( z ) \approx 
	- \frac{ \hbar \alpha(0) \Omega }{ 8 z^3 } 
	\frac{ \epsZero - 1 }{ \epsZero + 1 }
	\label{eq:vdW-local-dielectric}
\end{equation}
while the local Drude conductor gives \cite{Wylie1984}
\begin{equation}
	\mbox{local metal}, z \ll \lambda_A : \quad
	U( z ) \approx 
	- \frac{ \hbar \alpha(0) \Omega }{ 8 z^3 }
	\frac{ \omega_s }{ \Omega + \omega_s }
	\label{eq:vdW-local}
\end{equation}
Here, $\omega_s^2 = 2\pi \sigma / \tau$ is the surface plasmon frequency,
and we have neglected Ohmic losses
(i.e., $\omega_s, \Omega \gg 1/\tau$).
The composite surface models with their nonlocal response
give a van der Waals interaction that crosses over between these two
limits (see Fig.\ref{fig:CP-semiconductor} below). 
This is similar to what 
has been analyzed at large distances by Pitaevskii \cite{Pitaevskii08}. 

\begin{figure}[bth]
\includegraphics*[width=0.45\textwidth]{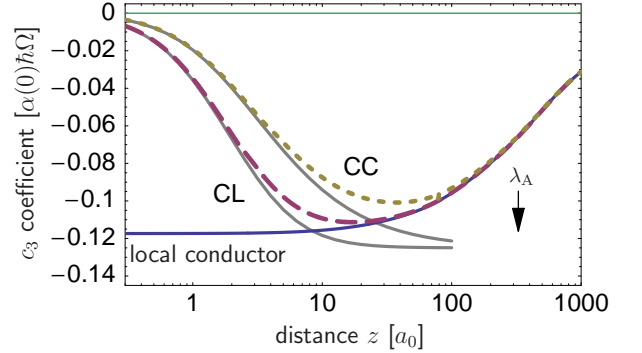}
% if figure file is missing
\caption[Energy-metal]{%
Atom-surface interaction at zero temperature,
from the van der Waals into the Casimir-Polder range.
We plot the $c_3$ coefficient, i.e., the 
potential $U( z )$ multiplied with $z^3$, in units of
$\hbar\Omega \alpha( 0 )$. The distance $z$ is in units of the
screening length $a_0$. The arrow marks the (reduced) transition
wavelength $\lambda_A$.
Solid (blue) line ``local conductor'': 
Drude metal with dielectric function~(\ref{local_dielectric}), 
dashed: surface covered with a diffusive charge layer (CL model), 
dotted: continuous bulk charge with a hydrodynamic response (CC model).
Solid gray lines: short-distance asymptotes to the CL and CC models,
based on the approximate reflection 
coefficients~(\ref{eq:large-k-limit-CL}, \ref{eq:large-k-limit-CC}).
\\
Parameters: DC conductivity $\sigma = 3.6\cdot 10^{17}\,{\rm s}^{-1}$
(typical for Al), dielectric constant $\epsZero = 1$,
electron scattering times $\tau = \tau_s = 10^{-15}\,{\rm s}$,
diffusion constants $D = D_s = 5\cdot10^3\, {\rm cm^2/s}$,
atomic resonance wavelength $2\pi \lambda_A = 628\,{\rm nm}$
($\Omega/2\pi = 477\,{\rm THz}$). 
For these para\-meters, the screening length is $a_0 \approx 0.3\,{\rm nm}$.
The van der Waals interaction $c_3 / z^3$ with a local Drude conductor
gives a normalized value $-0.118$ from Eq.(\ref{eq:vdW-local}).
The CC/CL models show, at short distances, a much weaker interaction
potential ${\cal O}( 1/ z )$ given in Eq.(\ref{eq:short-distance-vdW-CL}).
}
\label{fig:potential-metal}
\end{figure}

We focus here on the somewhat academic
case of a simple free-electron metal ($\epsZero = 1$ ), where the nonlocal
surface reponse changes even the exponent at short distances because
the $c_3$ coefficient in Eq.(\ref{eq:vdW-local-dielectric}) vanishes.
To get insight into the
small-distance behaviour, we expand
the CL reflection coefficient $r_{\rm TM}$ at high momentum and
get for $\tau_s = \tau$
%{\color{magenta}
%approximation in $v$: neglect
%$(\xi/c)^2 \epsilon( {\rm i} \xi )$ relative to $k^2$. For the good
%conductor where the conductivity term dominates, this implies the
%ratio $z^2 \xi 4\pi \sigma / c^2$ to be small. In terms of an
%electromagnetic diffusion coefficient $D_m = c^2 / 4\pi \sigma$:
%$z^2 \xi / D \ll 1$.} 
\begin{equation}
	%\mbox{CL},  
	k \sim 1/a_0 \gg \xi/c : \quad
	% r_{\rm TM}( {\rm i} \xi, k ) 
	\frac{ \epsilon_s( {\rm i}\xi, k ) k - v 
		}{ \epsilon_s( {\rm i}\xi, k ) k + v }
	\approx \frac{ 1 }{
	1 + 2 (k a_0)^2 }
	\label{eq:large-k-limit-CL}
\end{equation}
Note that the imaginary frequency $\xi$ drops out in this case and
the suppression $\sim 1/k^2$ on scales shorter than the screening length
$a_0$.
The $\xi$ integral in Eq.(\ref{eq:vdW-integral-xi-CL}) can then be performed, 
and one gets for $z \ll a_0$ the simple result
% {\color{magenta}[CH:] is anybody voting for writing down the result 
% for $\tau \ne \tau_s$?}
\begin{equation}
	\mbox{CL}, z \ll a_0 : \quad
	U( z ) \approx - \frac{ \hbar \Omega \alpha(0) }{ 8 z a_0^2 }
	\label{eq:short-distance-vdW-CL}
\end{equation}
instead of the $1/z^3$ power law. In Fig.\ref{fig:potential-metal},
we compare the exact evaluation of the van der Waals 
potential~(\ref{eq:CP-zero-T-imag-freq}) (dashed curve) 
to the numerical integration over
$k$ of the approximate reflection
amplitude~(\ref{eq:large-k-limit-CL}) (solid gray curve). One gets
a good approximation over a wide range of non-retarded
distances $0.1\, a_0 < z \ll \lambda_A$. Note how the non-local theory
matches with the local conductor (solid blue curve) as $a_0 \ll z$.

A similar analysis for the CC model [Table~\ref{tab:reflection coefficients}] 
yields a screening length $a_0 =\sqrt{\epsZero D/4\pi \sigma}$ involving 
the bulk diffusion constant, and the approximate form
\begin{equation}
	%\mbox{CL},  
	k \sim 1/a_0 \gg \xi/c : \quad
	r_{\rm TM}( {\rm i} \xi, k ) 
	% \frac{ \epsilon_s( {\rm i}\xi, k ) k - v 
	%	}{ \epsilon_s( {\rm i}\xi, k ) k + v }
	\approx 
	\frac{ \sqrt{ k^2 + 1/a_0^{2} } - k }{ \sqrt{ k^2 + 1/a_0^{2} } + k }
	\label{eq:large-k-limit-CC}
\end{equation}
% Fig.\ref{fig:potential-metal} shows the crossover
% around $z = a_0$
% from the van der Waals limit to a $1/z$ power law. 
Despite the difference in the reflection amplitudes~(\ref{eq:large-k-limit-CL}, 
\ref{eq:large-k-limit-CC}), the short-distance asymptote turns out to 
be just a factor of one half smaller than Eq.(\ref{eq:short-distance-vdW-CL})
(upper solid gray curve in Fig.\ref{fig:potential-metal}).
These results illustrate the dramatic impact of nonlocal electrodynamics on
the van der Waals interaction. We expect them to apply qualitatively
in materials where the background polarizability $\epsZero > 1$ provided
by bound electrons is relatively small. The $c_3$ coefficient
in Eq.(\ref{eq:vdW-local-dielectric}) is then nonzero, but weak and shifts 
the short-distance
asymptotes of the CL and CC curves in Fig.\ref{fig:potential-metal} below zero.

\section{Casimir-Polder (retarded) regime}
\label{s:Casimir-Polder}
This regime corresponds to an intermediate range of distances,
\begin{equation}
	\lambda_A \ll z \ll \lambda_T %\qquad
	\label{eq:Casimir-Polder-limit}
\end{equation}
where the thermal wavelength $\lambda_T %= \hbar c / k_{\rm B} T 
= 7.6\,\mu{\rm m}$ at $300\,{\rm K}$. This range is
characterized by the form ${\cal F}\sim 1/z^4$ for both dielectric and
metallic surfaces with a local dielectric response.

For analytic expansions, we consider $T=0$ so that the $\xi$-integral of 
Eq.(\ref{eq:U-CP-zero-temp}) still applies. From the exponent $2 v_0 z$,
we read off the characteristic frequency $\omega_c = c / (2 z)$ that
limits the $\xi$-integration range to $\xi 
\,\raisebox{-0.5ex}{$\stackrel{<}{\sim}$} \,\omega_c$. In the CP 
regime~(\ref{eq:Casimir-Polder-limit}),
$\omega_c$ is much smaller than the atomic resonance $\Omega$, 
and we can 
%Switching
%from $(\xi, k)$ to the integration variables
%$x \equiv \xi / \omega_c$ and
%$y \equiv c v_0 / \omega_c$, one gets the representation
%\begin{eqnarray}
%	U( z ) &=& - \frac{ \hbar c }{ 32 \pi \cdot z ^4}
%	\int^{\infty}_{0}\!{\rm d}x \, \alpha(i\omega_c x)
%	\int^{\infty}_{x} \!{\rm d}y \, e^{-y} \cdot {}
%	\label{eq:}
%\nonumber\\
%	&&	
%	{} \cdot \big[
%	(2y^2 -x^2)
%\frac{ \epsilon_s y - \sqrt{y^2 + x^2(\epsilon-1)}
%	}{ \epsilon_sy+\sqrt{y^2+x^2(\epsilon-1)} }
%\nonumber\\
%	&& \quad {}
%- x^2 \frac{ y - \sqrt{y^2 + x^2(\epsilon-1)}
%	}{ y + \sqrt{y^2 + x^2(\epsilon-1)} }
%	\big]	
%	\label{eq:CP-range-xy-integral}
%\end{eqnarray}
%Anticipating that the $y$-integral confines the relevant values for $x$
%to the range $0 \le x \leq 1$, we identify the small parameter
%$\omega_c / \Omega \ll 1$ and 
expand the polarizability
\begin{equation}
%	\alpha (i\omega_{c} x) =
%	\frac{\alpha (0)}{ 1 + x^2 (\omega_c/\Omega)^2 }
%	\approx \alpha (0)( 1 - x^2 (\omega_c/\Omega)^2 )
	\alpha ( {\rm i} \xi )
%	\frac{\alpha (0)}{ 1 + x^2 (\omega_c/\Omega)^2 }
	\approx \alpha (0)( 1 - \xi^2 / \Omega^2 )
	\label{eq:expand-alpha}
\end{equation}
The following discussion applies to a free-electron metal at high
density where $\epsZero = 1$ and $\sigma \gg \Omega$. 

\subsection{Good conductor}

We then have the small parameter $\omega_c / \sigma$ to
simplify the reflection coefficients. It turns out that the impact of
charge diffusion is very small in the Casimir-Polder regime: 
for the CL model, we find by inspection that the relevant
dimensionless ratio is $D_s / (c z ) \ll 1$. 
%
%There are then two small parameters in 
%Eq.(\ref{eq:CP-range-xy-integral}).
%In the reflection amplitudes, we observe the combination ($\epsZero = 1$
%here)
%\begin{equation}
%	x^2 ( \epsilon - 1 ) = \frac{ 2 x }{ a^2 ( 1 + x \tilde \tau ) }
%	\label{eq:magnetic-diffusion-term}
%\end{equation}
%with $\tilde{\tau}=\tau\omega_c$ and
%the small parameter
%\begin{equation}
%	a \equiv \sqrt{\frac{\omega_c}{2\pi\sigma }}\ll
%	\sqrt{\frac{\Omega}{4\pi\sigma}} \ll 1
%	\label{eq:identify-a-small}
%\end{equation}
%Finally, the effective surface permittivity involves for the CL model
%a small parameter $D_0 a^2$
%%	\be
%%	\epsilon  = 1+\frac{2\cdot 4\pi\sigma\cdot z}{c\cdot x(1+\tilde{\tau}x)}=1+ \frac{2}{x\cdot a^2(1+\tilde{\tau}x)}
%%\ee
%\begin{equation}
%	\epsilon_s = \frac{\epsilon x +
%	{\displaystyle \frac{ D_0 a^2(y^2-x^2)}{(1+\tilde{\tau}_sx)}}
%	}{ x + {\displaystyle \frac{D_0 a^2(y^2-x^2)}{(1+\tilde{\tau}_sx)}}
%	}
%	\label{eq:tilde-epsilon-dimless}
%\end{equation}
%defined by
%$D_0 \equiv 2\pi \sigma D_s / c^2$
%and
%$\tilde{\tau}_s = \tau_s\omega_c$.
We start with the zeroth order with respect to this ratio and
expand in powers of 
\begin{equation}
	\delta = \sqrt{ \frac{ \omega_c }{ 2\pi\sigma } } \ll 1
	\label{eq:identify-small-a}
\end{equation}
Performing the integrations, we get
the familiar Casimir-Polder potential and next-order corrections
\begin{eqnarray}
	U( z ) &\approx&
	-\frac{3\hbar c\alpha(0)}{8\pi z^4}
	\Big[ 1 - \frac{20}{3}\frac{\omega_c^2}{\Omega^2}
	+ \frac{ \delta }{ 6 \sqrt{2} } \int_0^\infty \!{\rm d}x
\sqrt{1 + \omega_c \tau \, x} \cdot {}
\nonumber
\\
&& \cdot \left(
	x^{7/2}\Gamma(0,x) - 3(1+x)e^{-x}x^{3/2}
\right)
	+ {\cal O}(a^2)
	\Big]
	\label{eq:expansion-CP}
\end{eqnarray}
where the incomplete Gamma function is
$\Gamma(0,x) \equiv \int_x^\infty t^{-1}e^{-t}dt$.
The limiting values of third term 
give a correction
$- (77 / 72) \delta \sqrt{ \pi / 2 } \approx - 1.34\,\delta
\sim z^{-1/2}$
as $\omega_c \tau \to 0$ and
$- (8 / 5) \delta \sqrt{ 2 \omega_c \tau } \approx - 2.26\,\delta \sqrt{ \omega_c \tau }
\sim z^{-1}$
as $\omega_c \tau \to \infty$.
%{\color{magenta}[* Elad, please check the numerical coefficients in
%Eq.(eq:expansion-CP), as my Mathematica integration does not recover
%your numerical result (missing factor $1/12$) *]
%}

Note that these approximations correspond to two inequivalent ways of
implementing the perfect-conductor limit. The first case could be called
``overdamped'', with a purely real conductivity. The penetration of
transverse fields into the bulk then occurs by means of diffusion.
The correction to the Casimir-Polder potential
in Eq.(\ref{eq:expansion-CP}) scales like $\ell_c / z \propto z^{-1/2}$ 
where $\ell_c =
\sqrt{c^2 / \sigma \omega_c}$ is the magnetic diffusion
length at the characteristic frequency $\omega_c$.
In the second case, the conductivity is purely imaginary and, 
similar to a superconductor, the transverse field is screened from
the bulk. The correction to the Casimir-Polder potential arises from
the field penetrating a thin layer of the order of the plasma
wavelength (also called London-Meissner penetration depth) 
$\lambda_p = c \sqrt{ \tau / (4\pi\sigma) }$,
and scales like 
$\lambda_p / z$. The latter case has been studied, for example,
in Ref.\cite{Bezerra08}, Eq.(37), and their result is recovered 
by the two correction terms in Eq.(\ref{eq:expansion-CP}):
\begin{equation}
	\frac{ U( z ) - U_{\rm CP}( z ) }{ U_{\rm CP}( z ) } \approx
	- \frac{5}{3}\frac{ \lambda_A^2  }{z^2 }
	- \frac{8}{5}\frac{\lambda_p }{ z } 
	% + {\cal O}[a^2]
	\quad ( \lambda_A \ll z \ll \tau c )
	\label{eq:relative-CP-correction}
\end{equation}
where $U_{\rm CP}( z )$ is the first term in Eq.(\ref{eq:expansion-CP}).
This range of distances is quite narrow for the parameters of 
Fig.\ref{fig:potential-metal}, where $c \tau \approx 300\,{\rm nm}$.

Let us now extract the contribution due to the diffusive charge layer
(CL model). To the first order in the surface diffusion coefficient
$D_s$, the correction to the local model can be worked out to be:
\begin{eqnarray}
	&&\frac{ U( z ) - U_{\rm loc}( z )}{ U_{\rm CP}( z ) }
	\approx - \frac{ D_s \delta }{ 6 \sqrt{ 2 } \, c z}
	\int^{\infty}_{0}\!{\rm d}x\,
	\frac{ \sqrt{ x } \, (1+\omega_c \tau \,x)^{1/2} }{
		1+ \omega_c \tau_s \,x }
	\cdot {}
\nonumber\\
	&& \quad {} \cdot
% e^{-x}(12 + x (12 - (x - 3) x))
	\left[
	e^{-x}(12 + 12 x + 3 x^2 - x^3 )
	+ x^4\Gamma(0,x)
	\right]
	\label{eq:CL-correction-to-CP}
\end{eqnarray}
In the limiting case $\omega_c \tau$, $\omega_c \tau_s \to 0$, 
the integration~gives a relative correction
$ - (285/88) \sqrt{ \pi / 2 } \, (D_s \delta / cz)
\sim z^{-3/2}$; in the opposite limit,
$ - (12/5) \sqrt{ 2 \tau / \omega_c \tau_s^2 } \, (D_s \delta / cz)
= - (6\sqrt{2}/5) (c_s/ c)^2 (\lambda_p / z)$ where the speed of
sound
$c_s = \sqrt{D_s / \tau_s}$ characterizes the dispersion of
longitudinal modes in the charge layer. The latter estimate illustrates
that the correction brought about by the charge layer is negligible
in the Casimir-Polder regime. A similar conclusion is reached for the
continuous charge (CC) model; we omit the calculations for brevity. 
The numerical results from Fig.\ref{fig:potential-metal} illustrate that
both CL and CC models merge into the local description in the Casimir-Polder
range 
$z \,\raisebox{-0.5ex}{$\stackrel{>}{\sim}$}\, \lambda_A$.
% (12/5) (1/\sqrt{2}) \lambda_p c_s^2 / c^2 z

% \subsection{Lifshitz (thermal) regime}

% need exponential exp( - 2 z v_0( i \xi, k ) ) for k = 0 and first Matsubara
% \xi = 2 \pi T:
% exp( - 4 \pi z T / c ) = \exp( - 4 \pi z / \lambda_T )
% We get the local result at a very high accuracy.

\subsection{Semiconductor}

\begin{figure}[h]
\includegraphics*[width=0.45\textwidth]{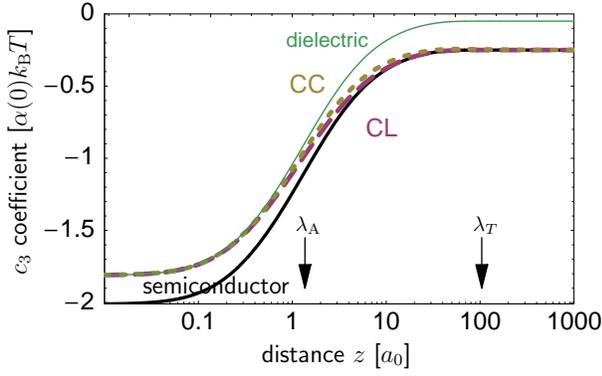}
\centering
\caption[Free energy]%
{Atom-surface potential through the Casimir-Polder range up to the
thermal wavelength. We plot
the $c_3$ coefficient of the free energy, i.e.
$\mathcal{F}( z ) z^3$, but here in units of $\alpha(0) k_{\rm B} T$. The
distance is normalized to the screening length $a_0 \approx 73\,{\rm nm}$.
Solid thick curve: local dielectric function in Drude form, dashed curve: 
charge layer (CL) model, dotted curve: hydrodynamic (continuous charge, 
CC) model, solid thin curve: non-conducting local dielectric.
\\
Parameters: background dielectric constant $\epsZero = 1.5$,
DC conductivity $\sigma = 10^{10}\,{\rm s}^{-1}$ (comparable to Ge),
electron scattering time $\tau = \tau_s = 10^{-13}\,{\rm s}$,
diffusion constants $D = D_s = 4.5\, {\rm cm^2/s}$,
atomic resonance $\Omega/2\pi = 477\,{\rm THz}$ (wavelength
$2\pi\lambda_A = 628\,{\rm nm}$), Temperature $T= 300\,{\rm K}$
(thermal wavelength $\lambda_T = 7.6\,\mu{\rm m}$).
}
	\label{fig:CP-semiconductor}
\end{figure}

Fig.\ref{fig:CP-semiconductor} 
shows numerical calculations of the interaction potential
for a material with a conductivity typical
for semiconductors. The characteristic length $a_0$ for screening is
then much larger and falls into the Casimir-Polder range of the atom-surface
potential. The data show that the CC and CL models interpolate between 
the limiting cases of a local Drude conductor and a non-conducting 
dielectric (where $\epsilon( \omega ) = \epsZero$ does not diverge at
zero frequency). 
We have taken the relatively low value $\epsZero = 1.5$ to amplify
the difference between the dielectric and the conductor in the local limit: 
their difference scales with $2/(\epsZero + 1)$.

In Fig.\ref{fig:CP-semiconductor}, we show the free energy of
interaction ${\cal F}( z )$ calculated
from the Matsubara sum~(\ref{eq:Wylie-Sipe-finite-T}).
At distances beyond the thermal wavelength $\lambda_T$,
the free energy follows a $1/z^3$ power law with a $c_3$ coefficient 
proportional to $T$ that we discuss in the following section. The difference
between dielectric and Drude conductor arises, for these parameters,
from the zeroth term in the Matsubara sum. This term is discussed in more
detail in Sec.\ref{s:zero-freq-term}. Indeed, in the other terms, 
the conductivity enters only in the ratio
$4\pi\sigma / \xi_l = 2 \hbar \sigma / (l k_{\rm B} T)$.
At room temperature, this ratio can be neglected 
compared to the background dielectric constant $\epsZero$
provided the conductivity
%In this limit, the only distinct term between the different models in the Matsubara summation is the $l=0$ term, since for $l\geq 1$ the terms are identical with $\epsilon_s\approx \epsilon \approx \epsZero$. At room temperature (T=300K) this limit takes the form 
$\sigma \ll 4\cdot 10^{13}\,{\rm s}^{-1}$.
This regime applies to a wide range of doped semi-conductors.

The van der Waals regime for this material is not described by 
Eq.(\ref{eq:vdW-local}) due to the low conductivity. Ignoring conductivity
completely, Eq.(\ref{eq:vdW-local-dielectric}) for a local dielectric gives
a short-range coefficient $c_3$ with a value $- 1.91\, \alpha( 0 ) k_B T$ 
in the units of Fig.\ref{fig:CP-semiconductor}: this corresponds
well to the full calculation.
We have checked that the small difference is actually due to relatively 
large deviations from the non-retarded approximation that was applied
to derive
Eq.(\ref{eq:vdW-local-dielectric}). A similar situation occurred
in Ref.\cite{Bostroem12} which discusses the Casimir force between 
two plates separated by a dielectric liquid.

\section{Lifshitz (thermal) regime}
\label{s:zero-freq-term}
%
%
%\begin{figure}[h]
%\includegraphics[width=0.5\textwidth]{\figpath F-SC2.pdf}
%\centering
%\caption[Free energy]%
%{CP-Free energy: $\frac{\mathcal{F}z^3}{\alpha(0)k_{\rm B} T}$ plotted as a function of $\frac{z}{a_0}$. Solid thick curves: local model, dashed curves: charge layer (CL) model,
%dotted curves: continuous bulk charge (CC) model, solid thin curves: ideal dielectric model.
%Parameters: DC conductivity $\sigma = 10^{2}\,{\rm s}^{-1}$,
%electron scattering time $\tau = \tau_s = 10^{-13}\,{\rm s}$,
%diffusion constants $D = D_s = 4.5\, {\rm cm^2/s}$,
%atomic resonance $\Omega/2\pi = 477\,{\rm THz}$ (wavelength
%$628\,{\rm nm}$),
%Temperature $T= 300\,{\rm K}$.
%}
%\end{figure}

This section deals with the long distance regime
\(
	\lambda_T \ll z
\)
where the leading contribution to the atom-surface potential
is given by the $l = 0$ term in the Matsubara sum~(\ref{eq:Wylie-Sipe-finite-T}). The other terms are proportional to the exponentially small factor
$\exp( - 4\pi l z / \lambda_T )$ and can be neglected if the $l = 0$ term
is nonzero. A glance at Fig.\ref{fig:CP-semiconductor} illustrates that the
thermal regime is already well borne out at $z \sim \lambda_T$ due to
the factor $4\pi$ in the exponential.

The static term in the Matsubara sum has been the subject of much
discussion in the field of dispersion interactions
% CH 19 Jun 12 -- citation added
\cite{Geyer2005,Klimchitskaya2009}.
To illustrate this, we
give the limiting forms of the free energy in the thermal range for an
ideal dielectric material
\be
\lambda_T \ll z: \qquad
\mathcal{F}( z ) \approx -\frac{ \alpha(0) k_{\rm B} T }{4 z^3} 
\frac{\epsZero -1}{\epsZero +1}
	\label{eq:thermal_dielectric}
\ee
while for a conductor in the same limit
\be
\mathcal{F}( z ) \approx -\frac{ \alpha(0) k_{\rm B} T }{ 4 z^3 }
	\label{eq:thermal_conductor}
\ee
In fact, the latter result is obtained for any material with a nonzero 
conductivity: as $\sigma \rightarrow 0$, the former (dielectric) result
is not obtained in a continuous manner
% CH 19 Jun 12 -- reference added
 \cite{Klimchitskaya08}. 
This is due to the static reflection
coefficient $r_{\rm TM}(0, k)$ which is equal to $1$ for any nonzero $\sigma$,
while setting $\sigma = 0$ from the start for a pure dielectric, one gets
$r_{\rm TM}(0,k) = (\epsZero - 1)/(\epsZero + 1)$. This difference between
conductor and dielectric is also visible in the Casimir-Polder range shown 
in Fig.\ref{fig:CP-semiconductor}. The discontinuity disappears only in the
limit $T = 0$ 
% CH 19 Jun 12 -- half-sentence added.
for the material parameters considered here.
% although for the other terms in the
% Matsubara sum the limit $\sigma \rightarrow 0$ is continuous. It is only
% in the limit $T = 0$ that the discontinuity disappears. 

This effect is actually an artefact of the description in terms of a local
material response (conductivity, dielectric function). Using a hydrodynamic
model similar to our CC, Pitaevskii has shown that the free energy shows
a continuous cross-over between the limiting cases 
Eqs.(\ref{eq:thermal_dielectric}, \ref{eq:thermal_conductor}). We show
now that the same is true for both CC and CL models considered here.

\begin{figure}[h]
\centering
\includegraphics*[width=0.45\textwidth]{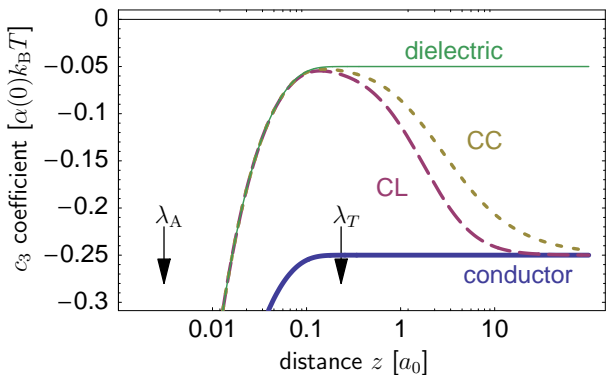}
\caption[Thermal range free energy]%
{Interaction potential in the thermal range: $c_3$ coefficient of the
free energy, i.e., $\mathcal{F}( z ) z^3$, in units of $\alpha(0) k_{\rm B} T$.
The distance is normalized to the screening length
$a_0 \approx 33\,\mu{\rm m}$.
Solid thick (blue) curve: local conductor 
($\sigma = 5 \cdot 10^{4}\,{\rm s}^{-1}$, typical for highly purified water), 
dashed curve: charge layer (CL) model with diffusion coefficient
$D_s = 4.5\, {\rm cm^2/s}$, $\tau_s = 10^{-13}\,{\rm s}$,
dotted curve: continuous bulk charge (CC) model ($D = D_s$, $\tau = \tau_s$),
solid thin (green) curve: ideal dielectric model ($\epsZero = 1.5$).
The other parameters are as in Fig.\ref{fig:potential-metal}.
%Parameters: DC conductivity 
%electron scattering time $\tau = \tau_s = 10^{-13}\,{\rm s}$,
%diffusion constants $D = D_s = 4.5\, {\rm cm^2/s}$
%(screening length $a_0 = 730\,\mu{\rm m}$).
% atomic resonance $\Omega/2\pi = 477\,{\rm THz}$ (wavelength
% $628\,{\rm nm}$),
% Temperature $T= 300\,{\rm K}$.
}
	\label{fig:c3-thermal-range}
\end{figure}

For the CC model, the first line of Eq.(\ref{eq:Wylie-Sipe-finite-T}) can
be written in terms of a dimensionless integral ($t = 2kz$)
\begin{eqnarray}
&&
\lambda_T \ll z: \qquad
	\label{eq:thermal-regime-CC}
\\
&& \mathcal{F}( z ) \approx -\frac{ \alpha(0) k_{\rm B} T }{ 8 z^3 } 
\int\limits_0^\infty \! dt\, t^2 e^{-t} 
	\frac{ \epsZero\sqrt{t^2 + (2z/a_0)^2} - t 
		}{ \epsZero\sqrt{t^2 + (2z/a_0)^2} + t}
\nonumber
\end{eqnarray}
with the screening length $a_0$ of Eq.(\ref{eq:def-a0}).
% $  R= \frac{z^2}{a_0^2} = \frac{z^2 4\pi \sigma}{\epsZero D}$
%
We recover Pitaevskii's result  \cite{Pitaevskii08} by calculating $a_0$
from the diffusion coefficient $D \approx k_{\rm B}T \tau / m^*$
of a non-degenerate electron gas.
This leads to $a_0^{-2} = n \ell_{\rm B}/(4\pi)$ where $n$ is the carrier 
density in the conductor and $\ell_{\rm B}$ the Bjerrum length 
% $ =  k_B T \epsZero / e^2$
(i.e., the distance where the Coulomb 
energy between two electrons becomes comparable to the thermal energy:
$e^2 / (\epsZero \ell_{\rm B}) = k_{\rm B} T$).
A glance at Eq.(\ref{eq:thermal-regime-CC}) tells that the dielectric and
metallic values of the reflection coefficient are smoothly interpolated as the
ratio $(z / a_0)^2$ changes from zero to infinity. This is 
illustrated in Fig.\ref{fig:c3-thermal-range} (dotted line) where the coefficient
of the $1/z^3$ power law is plotted vs $z / a_0$. 

% $a_0^{-2} = 4\pi e^2 n / \epsZero k_{\rm B} T$
% along with $\sigma=\frac{e^2n\tau}{m^*}$ leading to 
% $R = \frac{4\pi ne^2z^2}{\epsZero k_{\rm B}T}$.\\

The same qualitative behaviour is found in the CL model where the free
energy takes the form
% 
% $t=2kz , R_s = \frac{z^2 4\pi \sigma}{\epsZero D_s}$
\begin{equation}
\mathcal{F}( z ) \approx -\frac{ \alpha(0) k_{\rm B} T }{ 8 z^3 } 
\int\limits_0^\infty \! dt \, t^2 e^{-t}
	\frac{ (\epsZero - 1) t^2 + \epsZero (2 z / a_0)^2 
		}{ (\epsZero + 1) t^2 + \epsZero (2 z / a_0)^2 }
	\label{eq:thermal-regime-CL}
\end{equation}
% In the non-local models we note that for R or $R_s \ll 1$ we recover the ideal 
% dielectric result, while for $R$ or $R_s \gg 1$ we approach the local model 
% result, as demonstrated in Fig. 4. 
This is shown in dashed in Fig.\ref{fig:c3-thermal-range}. The parameters
chosen here are for a very poor conductivity (low carrier density) where the
screening length $a_0$ is large enough to fall into the thermal range.
This applies to dilutely doped semiconductors, or to the thermally 
excited conduction band of an intrinsic semiconductor.
% CH 28 Jun 12: "conducting" replaced by "conduction"
%
% {\color{magenta} ** Elad, please check the numbers and give
% $a_0$ in the caption. **}

\begin{figure}[h]
\includegraphics*[width=0.45\textwidth]{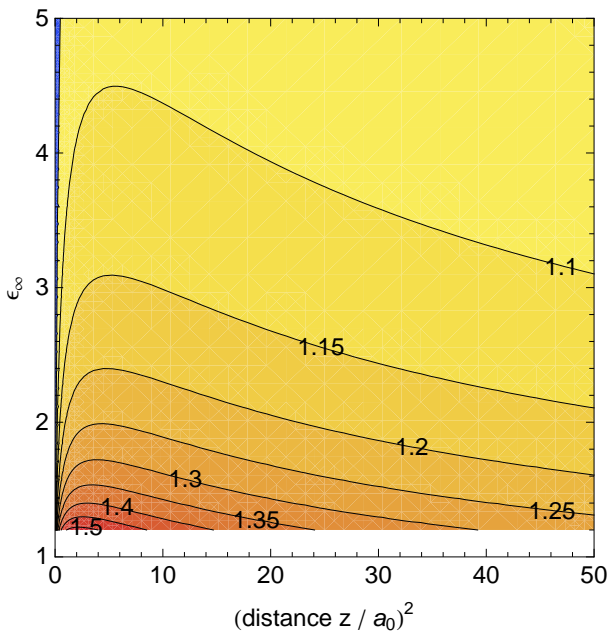}
\centering
\caption[Force difference between model -- CL vs CC]%
{Contour plot of the ratio between the interaction potentials in the 
CL and CC models. We consider only the thermal range 
Eqs.(\ref{eq:thermal-regime-CC}, \ref{eq:thermal-regime-CL}).
The $x$-axis gives the ratio $(z/a_0)^2$, the $y$-axis the
background dielectric constant $\epsZero$. The parameters for 
surface and bulk diffusion are the same ($D = D_s$, $\tau = \tau_s$).
%
% $\frac{f_{CL}}{f_{CC}}(R,\epsZero)$ (plotted as a function of R and 
% $\epsZero$), $R_s= R(D=D_s)$.
}
	\label{fig:ratio-CCCL-contour}
\end{figure}

% Fig. 5 shows the $l=0$ term, for which the CC/Cl difference is maximal. Since 
% the $l=0$ term is dominant in the thermal range we expect that the CC/CL 
% difference is maximal in this range. 
In Fig.\ref{fig:ratio-CCCL-contour}, we explore under which conditions the
atom-surface interaction energy is most sensitive to the details of the
surface charge response. This contour plot shows the ratio between
the CL and CC results for the interaction energy in the thermal range,
varying the distance $z$ and the background dielectric 
constant $\epsZero$. 
The two models differ maximally in the cross over range 
$z \approx a_0$, and for $\epsZero \sim 1$. This
could have been expected from Eqs.(\ref{eq:thermal_dielectric},
\ref{eq:thermal_conductor}) because the dielectric and metallic limits
(the two horizontal lines in Fig.\ref{fig:c3-thermal-range}) are then most
separated.

\section{Discussion and conclusion}
\label{s:discussion}

% A physical quantity
% that is particularly suited to extract this behaviour is the entropy of
% interaction, $S = - \partial F / \partial T$. 

Any material with mobile charges is characterized by a screening
length $a_0$, and a local current-field relation (Ohm's law) is necessarily limited 
to scales larger than $a_0$. We have explored in this paper how screening at and
below the surface influences the long-range van der Waals-Casimir-Polder 
(vdW-CP)
interaction between an atom and the body. Our description may be termed
`mesoscopic' in the sense that the electronic response is collective in nature,
but retains traces of ballistic carriers via the diffusion coefficient $D \sim
v \ell$ ($v$ is a typical carrier velocity and $\ell = v \tau$ the scattering
mean free path). Two models for the electromagnetic response of the
surface were studied in detail: a continuous charge distribution below the 
surface (CC) within a hydrodynamic approximation, and a thin charge layer (CL)
with a diffusive response typical for, e.g., localized surface states. 
Both models
provide a continuous crossover of the \vCP potential between two local
limiting cases: a pure dielectric and a conducting medium,
%
% CH 28 Jun 12: half-sentence added
as the atom-surface distance goes through the range $z \sim a_0$.
The two limits
can be distinguished from the zero-frequency limit of their dielectric functions
and give different coefficients $c_3$ for the $1/z^3$ power laws that prevail
at very short (van der Waals) and very large (thermal or Lifshitz) distances.
Our calculations extend the picture proposed in 
Refs.\cite{Pitaevskii08,Dalvit08} to any distances,
namely that the nonzero DC conductivity $\sigma$ can be neglected if the 
atom-surface distance $z$ is shorter than the screening length $a_0$
[Eq.(\ref{eq:def-a0})].

The differences in the vdW-CP interaction may be used as a probe
that can identify the type of charge transport in the (sub)surface region.
The sensitivity of this probe is maximal 
when the screening length $a_0$ matches the atom-surface distance $z$,
particularly in the retarded 
% ({\color{magenta}$\bullet$ `thermal' replaced by `retarded'}) 
range $z > \lambda_A$. 
The CC/CL difference is particularly large when the dielectric constant
$\epsZero$ of bound carriers and background ions is close to $1$. The reason 
is that the jump $\epsZero - 1$ leads to a surface polarization charge that 
responds locally (not by diffusion) and is therefore masking the effect of either 
the CC or the CL.
For transitions in the 
optical visible range, these favorable conditions correspond to 
$z>1\,\mu{\rm m}$ which is indeed 
the achievable range of present experiments. A screening length
$a_0\approx 1\,\mu{\rm m}$ occurs in a dilute semiconductor, which 
at room temperature is a non-degenerate electron system with 
$a_0 = \sqrt{\epsZero k_B T / (4\pi n e^2)}$. Hence we require a
carrier density $n \approx 10^{12}\, {\rm cm}^{-3}\times\epsZero$. 
This density can be achieved by dilute doping, as for 
fused silica \cite{Pitaevskii08}. One may also work with intrinsic
semiconductors where the carriers are thermally excited with a density
$n\approx 10^{19}\,{\rm cm}^{-3} \exp{(-E_g/2k_B T)}$
\cite{Dalvit08,ashcroft}
where $E_g$ is the gap. Hence, with $E_g\approx 1\,
{\rm eV}$ and by varying the temperature, one can span a range around
$10^{12}\,{\rm cm}^{-3}$.

In conclusion, our two charge type models are representatives of a composite surface of a metal, with a nonlocal electromagnetic response due to 
charge transport in the bulk and at the surface. Such composite surfaces are
fairly common corresponding to either surfaces covered with adsorbates or nanostructures, or to disordered surfaces with quantum well states. We have studied the Casimir-Polder interaction with such surfaces and have shown where this effect can be used as a sensitive probe of the surface type and its diffusive properties.

\smallskip
\noindent
\emph{Acknowledgements.}
This research was supported by a grant from the German-Israeli
Foundation for Scientific Research and Development (GIF). 
% E. Eizner thanks the University of Potsdam for hospitality.
We thank H. Haakh for useful discussions and G. L. Klimchitskaya and
V. M. Mostepanenko for comment.

% ================================================
%
%
%              Appendix
%
%
% ================================================

\appendix

\section{Surface response}
\label{a:calculation-TE-TM}

%\subsection{Diffusive currents and charge conservation}
%% \subsection{Diffusive charge layer}

%% jump in tangential magnetic field from layer current.

%In the charge layer model, 
%the response function for the current density ${\bf J}$ in the layer
%is
%\begin{equation}
%	{\bf J} = - D_s( \omega ) \nabla \gamma % = - {\rm i} k D \gamma
%	\label{eq:layer-current-response}
%\end{equation}
%where $D_s( \omega )$ 
%is the surface diffusion constant and $\gamma$ the surface
%charge density. We show in Sec.\ref{a:no-surface-conduction} that the
%conduction current in a thin layer is negligble. The surface current leads
%to a jump in the tangential magnetic field, see 
%Eq.(\ref{eq:By-jumped-outside}) below.
%%The second expression applies for an incident field
%The conservation of total charge (surface and bulk in $z < 0$)
%yields
%\begin{equation}
%	- {\rm i} \omega \gamma - D_s( \omega ) \nabla^2 \gamma =
%	j_z( 0^- )
%	\label{eq:charge-conservation}
%\end{equation}
%where $j_z( 0^- )$ is the normal component of the bulk current density,
%evaluated just below the charge layer. We focus in the charge layer model
%on a local bulk response so that $j_z( 0^- ) = \sigma( \omega ) E_z( 0^- )$ 
%with the conductivity $\sigma( \omega )$.

%%{\color{magenta}$\bullet$ a word on the semiconductor where one may
%%have some complex $p$-parameter multiplying the exciton = polarization
%%current $j_z( 0^- )
%%= - {\rm i} \omega (\varepsilon - \varepsilon_0 ) E_z( 0^- )$.
%%}

\subsection{Surface impedances}

% \subsection{Surface impedance formulation}

% ratio of tangential electric and magnetic fields, just outside the layer.
% links to reflection coefficient (separate formulas for the two polarizations).

%{\color{magenta}$\bullet$ use standard notation for impedances?
%I don't know whether there is a standard one...
%}

The calculation of the electromagnetic Green function proceeds by 
expanding the field created by a point source into Fourier components
and finding reflection and transmission coefficients for
each wave vector ${\bf k}_{\rm i}$ incident on the
surface. With the wave vector in the $xz$-plane and the macroscopic body
in the half-space $z < 0$, we have ${\bf k}_{\rm i}
= k {\bf e}_x - k_z {\bf e}_z$
with $k_z = \sqrt{ \omega^2 / c^2 - k^2 } = {\rm i} v_0$. We consider
separately two principal polarizations. In the TE-polarization,
the electric field outside the surface is written in the form
\begin{equation}
	z > 0: \quad
	{\bf E}( {\bf r} ) = E_{\rm TE}( k ) \, {\rm e}^{ {\rm i} k x }
	{\bf e}_y \left(
	{\rm e}^{ v_0 z } + r_{\rm TE} {\rm e}^{ - v_0 z }
	\right)
	\label{eq:TE-field-ansatz}
\end{equation}
where ${\bf e}_y$ is the unit vector transverse to the plane of incidence.
One gets the magnetic field from the Faraday equation: 
$B_x( {\bf r} ) = {\rm i}(c/\omega) \partial_z E_y( {\bf r} )$. The ratio
between these two tangential fields, evaluated at $z = 0^+$, is the
surface impedance $Z_{\rm TE}$ and determines the reflection coefficient
\begin{equation}
	Z_{\rm TE} = \frac{ E_y( 0^+ ) }{ B_x( 0^+ ) }, \qquad
	r_{\rm TE} = \frac{ {\rm i} (c v_0 / \omega ) Z_{\rm TE} - 1
		}{ {\rm i} (c v_0 / \omega ) Z_{\rm TE} + 1 }
	\label{eq:TE-r-and-Z}
\end{equation}
In a local model for the body response, one has $Z_{\rm TE}
= - {\rm i} \omega / (c v)$ where the transmitted wave vector is
${\bf k}_{\rm t} = k {\bf e}_x - {\rm i} v {\bf e}_z$, and the Fresnel
formula is recovered.

In the TM-polarization the electric field vector is in the $xz$-plane,
and Eq.(\ref{eq:TE-field-ansatz}) becomes
\begin{eqnarray}
	{\bf E}( {\bf r} ) &=& A_{\rm TM}( k ) \, {\rm e}^{ {\rm i} k x }
	\Big[ ({\rm i} k {\bf e}_z - v_0 {\bf e}_x)
	{\rm e}^{ v_0 z }
	\nonumber\\
	&& {}
	+
	r_{\rm TM}
	({\rm i} k {\bf e}_z + v_0 {\bf e}_x)
	{\rm e}^{ - v_0 z }
	\Big]
	.
	\label{eq:TM-field-ansatz}
\end{eqnarray}
This gives
$B_y = - {\rm i}(c/\omega) ( \partial_z E_x - {\rm i} k E_z) =
- {\rm i} (\omega / c) A ( 1 + r_{\rm TM} )$ just above the surface.
One defines impedance and reflection coefficient from the tangential
($x$-) component of the electric field
\begin{equation}
	Z_{\rm TM} = \frac{ E_x( 0^+ ) }{ B_y( 0^+ ) }, \qquad
	r_{\rm TM} = \frac{ v_0 - {\rm i} (\omega / c) Z_{\rm TM}
		}{ v_0 + {\rm i} (\omega / c) Z_{\rm TM} }
	\label{eq:TM-r-and-Z}
\end{equation}
Its local approximation is $Z_{\rm TM} = - {\rm i} c v / 
[\omega \epsilon( \omega )]$
where $\epsilon( \omega )$ is the bulk dielectric function.

\subsection{Solving the reflection problem}

We start to work out the electromagnetic response function in the
TE-polarization. Within a local description of the bulk medium below
the layer, one can work with a medium wave vector
${\bf k}_{\rm t}$ in the $xz$-plane, as defined after
Eq.(\ref{eq:TE-r-and-Z}), with
\begin{equation}
	{\rm i} v = \sqrt{ (\omega/c)^2 \epsilon( \omega ) - k^2 }
	\label{eq:def-v-in-medium}
\end{equation}
An ansatz similar to Eq.(\ref{eq:TE-field-ansatz}) can be written down
and augmented by a longitudinal part
\begin{eqnarray}
	z < 0: \quad
	{\bf E}( {\bf r} ) &=& E_{\rm TE}( k ) \, {\rm e}^{ {\rm i} k x }
	\Big[
	{\bf e}_y t_{\rm TE}	{\rm e}^{ v z }
\nonumber\\
	&& {} +
	t_{L} (
	{\rm i} k {\bf e}_x + v_1 {\bf e}_z )
	{\rm e}^{ v_1 z }
	\Big]	
	\label{eq:TE-field-inside}
\end{eqnarray}
where the component $v_1$ of the longitudinal wave vector is as yet
undetermined. From the Maxwell equations, the tangential component
$E_x$ is continuous, and since it is zero above the layer
{}[Eq.(\ref{eq:TE-field-ansatz})], we find $t_L = 0$ for the longitudinal
amplitude. The field $E_z$ perpendicular to the surface
is zero above and below the layer, hence the surface charge
$\gamma$ and the current density ${\bf J}$ are zero from
Eqs.(\ref{eq:diffusive-current-response}, \ref{eq:layer-bulk-charge-conservation}).
The magnetic
field $B_x$ is then continuous as well, and we get
the local value for the surface impedance from
\begin{equation}
	Z_{\rm TE} = \frac{ E_y( 0^- ) }{ B_x( 0^- ) } =
	\frac{ t_{\rm TE} }{ {\rm i} (c/\omega) v \, t_{\rm TE} }
	\label{eq:result-TE-impedance}
\end{equation}
The reflection coefficient~(\ref{eq:TE-r-and-Z}) takes the familiar form
\begin{equation}
	r_{\rm TE} = \frac{ v_0 - v }{ v_0 + v }
	\label{eq:result-TE}
\end{equation}
%{\color{green}
%$\bullet$ just for our understanding: if for some reason, the surface
%develops a magnetization layer (inhomogeneous current density below
%the surface), one would get $r_{\rm TE}$ that differs from Fresnel.
%}

In the TM-polarization, both transverse and longitudinal fields
in the medium are relevant, as is well known~\cite{Ford84}. The
expansion~(\ref{eq:TE-field-inside}) becomes
\begin{eqnarray}
	z < 0: \quad
	{\bf E}( {\bf r} ) &=& A_{\rm TM}( k ) \, {\rm e}^{ {\rm i} k x }
	\Big[
	({\rm i} k {\bf e}_z - v {\bf e}_x)
	t_{\rm TM} \, {\rm e}^{ v z }
	\nonumber\\
	&& {}
	+
	({\rm i} k {\bf e}_x + v_1 {\bf e}_z)
	t_{L} \, {\rm e}^{ v_1 z }
	\Big]
	.
	\label{eq:TM-field-inside}
\end{eqnarray}
The tangential field $E_x$ is continuous and becomes outside the layer
\begin{equation}
	E_x( 0^+ ) = A_{\rm TM}( k ) \, {\rm e}^{ {\rm i} k x }
	(
	- v t_{\rm TM} + {\rm i} k t_{L}
	)
	\label{eq:Ex-continued-outside}
\end{equation}
Due to the surface current density $J_x$, % = - {\rm i} k D_s \gamma$,
the field $B_y$ has a jump, and one gets above the layer:
\begin{eqnarray}
	B_y( 0^+ ) &=& B_y( 0^- ) - \frac{ 4\pi }{ c} J_x
\nonumber\\
	&=& - {\rm i} \frac{ \omega \epsilon( \omega ) }{ c } 
	A_{\rm TM} t_{\rm TM}
	+
	\frac{ 4\pi }{ c } {\rm i} k D_s( \omega ) \gamma	
	\label{eq:By-jumped-outside}
\end{eqnarray}
We need to express $\gamma$ and $t_{L}$ in terms of the transmitted
amplitude $t_{\rm TM}$: use charge
conservation~(\ref{eq:layer-bulk-charge-conservation}) and the continuity
of the $z$-component of the Amp\`ere-Maxwell equation which
links the jumps in $B_y$ (surface current) and in $\epsilon E_z$ (surface
charge). A straightforward calculation yields
\begin{equation}
	\gamma = - \frac{ \sigma( \omega ) k A_{\rm TM} t_{\rm TM} }{
	\omega + {\rm i} D_s( \omega ) k^2 }
	\label{eq:eliminate-gamma}
\end{equation}
and $t_{L} = 0$. Putting Eqs.(\ref{eq:Ex-continued-outside})--(%
\ref{eq:eliminate-gamma}) into
Eq.(\ref{eq:TM-r-and-Z}), we thus find the impedance
\begin{equation}
	Z_{\rm TM} =
	\frac{ - {\rm i} (c/\omega) v (1 + {\rm i} D_s( \omega ) k^2 / \omega) }{
	\epsilon( \omega ) %	\varepsilon_0 + 4\pi {\rm i} \sigma / \omega
	+
	{\rm i} \epsZero D_s( \omega ) k^2 / \omega
	}
%	\quad \mbox{(right)}
	\label{eq:result-Z-TM}
\end{equation}
%
%\begin{equation}
%	Z_{\rm TM} = \frac{ - {\rm i} c v ( \omega + {\rm i} D_s k^2 )
%	}{ \omega^2 \varepsilon_0 + 4\pi k^2 c D_s }
%	\quad \mbox{(wrong)}
%\end{equation}
where $\epsZero$ is the background dielectric function (excluding
the conduction current).
%{
%\color{magenta}$\bullet$
%check that this reproduces the TM-reflection coefficient found for the CL
%}

%in $zz$-response function, reflection coefficient
%\begin{eqnarray}
%	r_{zz} &=& \frac{ \epsilon_s v_0 - v }{ \epsilon_s v_0 + v },
%\\
%	\epsilon_s &=& \frac{ \epsZero
%	+ \frac{ 4\pi {\rm i} \sigma }{ \omega }
%	+ \frac{ {\rm i} \epsZero D_s k^2 }{ \omega }
%	}{ 1 + \frac{ {\rm i} D_s k^2 }{ \omega } }
%	\label{eq:Elad-rzz}
%\end{eqnarray}
%Hence
%\begin{eqnarray}
%%	{\rm i} (\omega/c) Z_{zz} = v / \epsilon_s
%	Z_{zz} &=& - {\rm i} \frac{ c v }{ \omega \epsilon_s }
%	\label{eq:}
%\end{eqnarray}

%

\subsection{Surface conductivity}
% \section{Conducting surface current}
\label{a:no-surface-conduction}

If we model the charge layer as a film of thickness $a$ and conductivity
$\sigma_s( \omega )$, its integrated current density (parallel to the 
layer) has the form:
\begin{equation}
\vec{J}({\bf{r}},\omega) 
=
% \sigma(\omega)\vec{E}({\bf{r}},\omega)\theta(-z)
- D_s(\omega) \vec{\nabla}_{||} \gamma(x,y) % \delta(z) 
%	\nonumber
%	\\
%&& 
+ \sigma_s(\omega) a \vec{E_{||}} % \delta(z)
\end{equation}
where the last term is the conduction current. Including this term in the 
surface response calculations we get reflection coefficients
\begin{eqnarray}
	r_{\rm TM} &=& \frac{\epsilon'_s v_0 - v}{\epsilon'_s v_0 + v}
	\label{eq:layer-TM}
\\
	r_{\rm TE} &=& \frac{v_0 - v + 4\pi i  a \sigma_s(\omega)\omega / c^2}{v_0 + v - 4\pi i  a \sigma_s(\omega)\omega / c^2}
	\label{eq:layer-TE}
\end{eqnarray}
with the surface dielectric function $\epsilon'_s$ being 
[cf. Eq.(\ref{eq:CL_dielectric})]
\begin{equation}
\epsilon'_s = 
\epsZero + 
4\pi i \frac{ \sigma(\omega) + v a \sigma_s(\omega)
}{ 
\omega + i D_s(\omega)k^2
}
	\label{eq:layer-surface-epsilon}
\end{equation}
%\epsilon'_s = \frac{\epsZero +\frac{i4\pi\sigma(\omega)}{\omega}+\frac{i\epsZero D_s(\omega)k^2}{\omega} + \frac{i4\pi va \sigma_s(\omega)}{\omega}}{1+\frac{iD_s(\omega)k^2}{\omega}}
We now identify under which conditions the terms proportional to
$\sigma_s(\omega)$ are negligible in these expressions.
For an order of magnitude estimate, we take
$\sigma_s(\omega) \approx \sigma(\omega)$ and 
a layer thickness $a$ the atomic scale.
%\be
%\tilde{\sigma_s}(\omega) = \sigma_s(\omega)\cdot a\approx \sigma(\omega)\cdot a
%\label{sigmas}
%\ee
Therefore we get from Eq.(\ref{eq:layer-surface-epsilon}) the condition
\begin{equation}
	v a
	= 
	a  \sqrt{ k^2 - \frac{\omega^2}{c^2}\epsilon(\omega) }
	\ll 1
	\label{eq:estimate-v-a}
\end{equation}
An upper limit can be found easily at imaginary frequencies $\omega = 
{\rm i} \xi$ where
$\epsilon(i\xi) = \epsZero + 4\pi\sigma / [\xi(1+\xi\tau) ] < 
\epsZero + \omega_p^2 / \xi^2$, with 
$\omega_p = \sqrt{4\pi\sigma/\tau}$ the plasma frequency.
Using the characteristic scales $k \sim 1/z$ and $\xi \sim \Omega$ in
the integrals for the Casimir-Polder potential, 
the estimate~(\ref{eq:estimate-v-a}) becomes
%
% and that the the dominant $k$ integration range is at 
% $k\approx \frac{1}{z}$, ${\tilde\sigma}_s(\omega)$ can be neglected 
% in $r_{TM}$ if
\be
a \sqrt{ \frac{1}{z^2} 
	+ \frac{\epsZero}{\lambda_A^2} 
	+ \frac{ 1 }{\lambda_p^2} 
	}
\ll 1
	\label{eq:layer-estimate-1}
\ee
Hence we require the layer thickness $a$ to be much smaller than
the smallest of the length scales distance $z$, atomic transition
wavelength $\lambda_A = c / \Omega \sqrt{\epsZero}$ and 
plasma wavelength $\lambda_p = c / \omega_p$.
All these are conditions are well satisfied for $a \le 1\,{\rm nm}$.
The surface dielectric functions $\epsilon'_s$ and $\epsilon_s$
[Eqs.(\ref{eq:layer-surface-epsilon}), (\ref{eq:CL_dielectric})] and
the $r_{\rm TM}$ reflection coefficients~Eq.(\ref{eq:layer-TM}),
Table~\ref{tab:reflection coefficients} are equivalent. 

As for the $r_{\rm TE}$ amplitude~(\ref{eq:layer-TE}), 
we can neglect the surface conductivity term provided
\begin{equation}
	\frac{ 4\pi a \sigma_s(i\xi)\xi }{ c^2 }
	\ll 
	\sqrt{ \frac{4\pi \sigma(i\xi) \xi }{c^2} } \le
	v
	\label{eq:rTE-unchanged}
\end{equation}
%\be
%\sqrt{\frac{\xi4\pi\sigma(i\xi)}{c^2}}\gg 
%\frac{4\pi\tilde{\sigma_s}(i\xi)\xi}{c^2}
%\ee
%% using (\ref{sigmas}) we get
As an estimate, this is equivalent to
\be
\frac{4\pi \sigma a^2 \xi}{c^2(1+\tau\xi)} \ll 1
\ee
The maximal value on the left hand side is 
$(a/ \lambda_p)^2$, hence $a$ must be smaller than the plasma 
wavelength, as we found before in~(\ref{eq:layer-estimate-1}).
%
% what about the limit $v_0 \sim v$? The difference in the local limit
% scales as
% \begin{equation}
%	v - v_0 \approx (\epsilon - 1 ) \xi^2 / 2 k c^2
% \approx 2\pi \sigma \xi / k c^2 (1 + \xi \tau)
%	\label{eq:}
%\end{equation}
%to be compared to
%\begin{equation}
%	 4\pi \sigma a \xi / c^2 (1 + \xi \tau)
%	\label{eq:}
%\end{equation}
%Hence $k \sim 1/z \gg a$ is sufficient: same condition as before. 
%
In the limit of large $k$, the difference $v_0 - v$ in Eq.(\ref{eq:layer-TE})
becomes small, and the condition~(\ref{eq:rTE-unchanged}) must be
replaced by
\begin{equation}
	\frac{ 4\pi a \sigma_s(i\xi) \xi }{ c^2 }
	\ll
	\frac{ 2\pi \sigma( i \xi ) \xi }{ k c^2 }
	\label{eq:}
\end{equation}
Estimating $k \sim 1/z$, we get $a \ll z$ which we also found 
in Eq.(\ref{eq:layer-estimate-1}) above.
For $a \le 1\,{\rm nm}$, also the $r_{\rm TE}$ amplitude is therefore
not affected by the layer conductivity.

% BibTeX users please use
%\newcommand{\mybstpath}{/Users/carstenh/Biblio/Database/bst/}
%\bibliographystyle{\mybstpath epj}

%\newcommand{\mybibpath}{/Users/carstenh/Work/Dropbox/CPlayer/}
%\bibliography{\mybibpath references}

%
%% Non-BibTeX users please use

\end{document}